\documentclass[12pt]{aa}
\usepackage[english]{babel}
\usepackage{graphicx}

\onecolumn
\newcommand{\ab}{Astrophys. Bull. }
\newcommand{\bsao}{Bull. Spec. Astrophys.Observ.}
\newcommand{\arep}{Astron. Rep. }
\newcommand{\alet}{Astron. Let. }
\newcommand{\araa}{Ann. Rev. Astron. Astrophys. }
\newcommand{\mnras}{Mon. Not. R. Astron. Soc. }
\newcommand{\apj}{Astrophys. J. }
\newcommand{\apjs}{Astrophys. J. Suppl.}
\newcommand{\aj}{Astron. J. }
\newcommand{\aaa}{Astron and Astrophys.}
\newcommand{\aas}{Astron and Astrophys. Suppl.}
\newcommand{\pasp}{Publ. Astron. Soc. Pasif. }

\begin{document}

\title{On the evolutionary status of  high-latitude \newline variable V534\,Lyr}
\author{V.G.~Klochkova, E.G.~Sendzikas and E.L.~Chentsov}

\institute{Special Astrophysical Observatory RAS, Nizhnij Arkhyz, 369167 Russia}
\date{\today} 

\abstract{Based on the high resolution spectral monitoring conducted at the 6-m
BTA telescope, we study the optical spectrum of the high-latitude variable V534\,Lyr. 
Heliocentric radial velocities $V_r$ corresponding to the positions of all metal 
absorption components, as well as the Na\,I D and H$\alpha$ lines were measured 
during  all the observational dates. The analysis of the velocity field examining the
lines of various nature revealed a low-amplitude variability of $V_r$ based on the 
lines with a high excitation potential, which are formed in  deep layers of the 
stellar atmosphere, and allowed to estimate the systemic velocity of  $V_{\rm sys} \approx -125$~km\,s$^{-1}$  
($V_{\rm lsr}\approx-105$~km\,s$^{-1}$). The distance estimate of  $d \approx 6$~kpc for 
the star leads us to its absolute magnitude of  $M_V \approx -5\fm3$, what corresponds 
to the MK spectral classification. The previously undetected  for this star spectral phenomenon 
was  revealed:  at certain times a splitting of the profiles of low-excited absorptions 
is observed, reaching ${\Delta V_r}=20$--$50$~km\,s$^{-1}$. A combination of the
parameters: reduced metallicity  ${\rm [Met/H]_\odot}=-0.28$, increased  nitrogen abundance
 ${\rm [N/Fe]}=+1.10$,  large spatial velocity, high luminosity, a strong 
variability of the emission-absorption profiles of H\,I lines, splitting of 
metal absorptions at different times of observations and the variability of 
the velocity field in the atmosphere allow  us to  classify V534\,Lyr as a 
pulsating star near the HB and belonging to the thick disk of our Galaxy.
 }

\authorrunning{\it Klochkova et al} \titlerunning{\it  On the evolutionary status of  high-latitude variable V534\,Lyr }

\maketitle

\section{Introduction}

During the past two decades, the 6-meter telescope BTA  hosts an investigation 
of the spectra of numerous  supergiants with a large excess of IR-radiation, presumably 
belonging to the  stage of the asymptotic giant branch (AGB)  and post-AGB.
The program also includes a study of some related stars with an unclear evolutionary status. 
At the post-AGB stage, far  evolved intermediate-mass stars are observed, with initial masses
at the main sequence  $2\div9\mathcal{M}_\odot$. Before the AGB
stage, these stars are cool (with  effective temperature of
\mbox{$T_{\rm eff} \approx 3000 $--$ 5000 $~K)} red supergiants.
After they have exhausted helium, a degenerate carbon-oxygen core is
formed in the core of \mbox{AGB stars} with an initial mass between
2--4\,$\mathcal{M}_\odot$. This core is surrounded by alternately energetically
active layers of He and H combustion. Most of the time,  the hydrogen
layer provides the energy release, but as the products of hydrogen
burning join the helium layer, a brief ignition of helium occurs in
the latter. This configuration of the internal structure of the star
is unstable, the theory predicts a fairly effective mixing and
dredge-up (due to the penetrating convection) of heavy metals
synthesized in nuclear reactions, accompanying these processes of
energy release into the atmosphere of the star (see the
survey~[\cite{Herwig}] and references therein).

Our initial goal was to determine the fundamental parameters of
investigated stars and  to  search for the chemical composition
anomalies in their atmospheres, related with the synthesis of
chemical elements during the previous evolutionary stages. As the
program runs, additional research became crucial, aimed to search for
the  time variability of spectral features and the velocity fields in
the atmospheres and envelopes of studied stars. This expansion of the
task required conducting for each program object of multiple
observations with high spectral resolution in a broad wavelength
range. The main results we have obtained from the spectroscopic data
at the \mbox{6-m} telescope are summarized in a number of recent
publications~[\cite{Envelopes,Enrich,Conf}]. Let us briefly note the
most important of them. Firstly, the parameters and chemical
composition of the atmospheres for several tens of stars with an
excess in infrared flux were determined. Secondly, evolutionary
variations in chemical composition (a large excess of the $s$-process
or hot-bottom-process elements) were found for  seven
stars~[\cite{04296,23304,Egg1,20000,20572}]. Based on the representative 
sample of stars, studied by us and the authors of~[\cite{Winckel}], we formed 
a subsample of post-AGB stars with the atmospheres rich in the $s$-process 
carbon and heavy metals, and with the carbon-rich circumstellar envelopes. 
The analysis of  properties of this subsample led us to the conclusion on 
the interrelation of the peculiarity of line profiles,  manifested in the presence 
of the emission component of the Na\,I doublet D lines, in the character of
molecular features, asymmetry, permitted or forbidden metal emissions, formed 
in the envelopes, and in the splitting of strong absorption profiles with 
a low lower-level excitation potential, with kinematic and chemical properties 
of the circumstellar envelope and with the type of its morphology~[\cite{Envelopes}].

The presence of the above-mentioned features is the main difference
of the spectra of post-AGB stars from the spectra of massive
supergiants. The splitting into the components of the profiles of
strongest heavy-metal absorptions in the spectra of selected post-AGB
supergiants we have detected allowed us to conclude that the process
of formation of a structured circumstellar envelope is accompanied
with an enrichment by the products of stellar
nucleosynthesis~[\cite{Enrich}]. Variability of the observed profiles
of the absorption-emission H$\alpha$ line  and metal lines, as well
as the type change (absorption/emission) of the C$_2$  molecule Swan
bands    registered in several objects are caused by the variations
of the parameters or the structure of the circumstellar envelope. The
H$\alpha$ profile type  (pure  absorption, pure  emission,  P\,Cyg or
inverse P\,Cyg-type, with two emission components in the wings) is not
  related to the chemical composition of the atmosphere of the central star.
The main factors influencing the  H$\alpha$ profile type and its
variability are the mass loss rate, the velocity of stellar wind,
kinematics and optical thickness of the envelope.

We have discovered the peculiarities, previously unknown in the
optical spectra of the AGB-candidate stars. In particular, based on
the observations, also performed with the NES  echelle spectrograph
of the 6-m telescope, the spectral features and the velocity fields
in the atmosphere and the envelope of the cool \mbox{AGB-supergiant}
V1027\,Cyg~[\cite{V1027Cyg}], the optical component of the \mbox{IR
source} IRAS\,20004+2955 were studied. For the first time, the
spectrum of this star revealed the splitting of the cores of strong
metal absorptions and their ions (Si\,II, Ni\,I, Ti\,I, Ti\,II,
Sc\,II, Cr\,I, Fe\,I, Fe\,II, Ba\,II). The  broad  profile of these
lines contains a stably located weak emission in the core, the
position of which can be considered as the velocity of the center of
mass of the system (for brevity, we shall further refer to it as the
systemic velocity) $V_{\rm sys}=5.5$~km\,s$^{-1}$. From the symmetric
small and moderate-intensity absorptions, weak radial velocity
variations  were revealed with an amplitude of
\mbox{5--6}~km\,s$^{-1}$, caused by pulsations. A red-wave H$\alpha$
profile shift was observed, caused by the distortion of the line
core. In the red spectral region, numerous weak lines of the CN
molecule and the K\,I\,7696\,\AA\ line, which has a P\,Cyg-type
profile are identified. An agreement of radial velocity, measured
from the symmetric absorptions of metals and the CN lines indicates
formation of a CN spectrum in the atmosphere of the star. Numerous
interstellar bands, DIBs, were identified, the position of which in
the spectrum, $V_r{\rm (DIBs)}=-12.0$~km\,s$^{-1}$, corresponds to
the velocity of the interstellar medium in the Local Arm of the
Galaxy.

Absolutely new results were obtained for a  nameless faint
star~[\cite{RAFGL5081}], associated with a powerful source of IR
radiation RAFGL\,5081. Its optical spectrum has been studied for the
first time based on the long-term spectral monitoring with high
spectral resolution. The estimates of the spectral class of the star
Sp\,$\approx$\,G5--8\,II and its effective temperature \mbox{$T_{\rm
eff}\approx 5400$~K} were made. We detected a hard-to-explain
spectral phenomenon: a splitting of the medium and low-intensity
profiles of stationary absorptions. Stationarity of absorptions
eliminates the possibility of explaining the double peaks owing to
the spectral binarity of the star. Radial velocities for the wind
components of the Na\,I D lines and H$\alpha$ line profiles reach
\mbox{$-250$}~km\,s$^{-1}$ and \mbox{$-600$}~km\,s$^{-1}$
respectively. These profiles contain narrow components, the number,
the depths and positions of which vary with time. The time-variable
multicomponent structure of the Na\,I D lines and H$\alpha$ line
profiles indicates the heterogeneity and instability of the
circumstellar envelope of RAFGL\,5081. According to the presence in
the Na\,I\,(1) line profiles of the components with the velocity of
\mbox{$V_r\rm{( IS)}=-65$}~km\,s$^{-1}$, it is concluded that
RAFGL\,5081 is located behind the Perseus Arm, i.e. not closer than
2~kpc. Note that this object is associated with a reflective nebula
GN\,02.44.7. The problem of photometric monitoring for determining
the variability parameters of the central star RAFGL\,5081 is set.

In the study of sources with a large IR flux excess, we identified
the stars outside the expected AGB and post-AGB stages. It turned out
that among such objects, in addition to AGB- and post-AGB stars,
there exist massive stars of extremely high luminosity with extended
and structured circumstellar envelopes. The most famous object that
has  for many years been preferably considered as a star in the
post-AGB stage is V1302\,Aql,  which has numerous peculiarities in
the spectrum. For a long time, the evolutionary status of this
supergiant, associated with a powerful IR source  IRC+10420, was
unclear. The set of observed properties of V1302\,Aql allowed to
consider it as a star at the post-AGB stage or as a very massive
star, past the red supergiant stage. It is obvious that depending on
the accepted status, and hence on the luminosity of the object too,
the estimate of its distance from the observer may differ several
times. One of the decisive arguments, confirming  the status of a
massive evolved star for V1302\,Aql was obtained by us in the
analysis of spectral data at the 6-m telescope when the chemical
composition of its atmosphere  was first determined and  a
significant excess of nitrogen was found~[\cite{IRC1}]. Detection of a
rapid growth of the effective temperature~[\cite{IRC1,Oudm98}] allowed
to assume that the star quickly evolves with the $T_{\rm eff}$
increase rate of about 120~K a year. The phenomenon was for us an
incentive to continue the spectral monitoring of this mysterious
object. In the study of the V1302\,Aql spectra over the period of
observations of 1992--2014  the conclusion was drawn that the yellow
hypergiant entered the phase of deceleration  (or the suspension) of
growth  of the effective temperature and approached at the H--R
diagram  the high-temperature boundary of the Yellow
Void~[\cite{IRC3}]. It is appropriate to mention here the star
HD\,179821, which has for several decades appeared in the papers as a
post-AGB candidate, and only a detailed analysis of the high-quality
data of  a long-term spectral monitoring with the echelle
spectrographs of the SAO RAS 6-m BTA telescope and the   McDonald
observatory 2.7-m telescope~(USA) led to the conclusion that it
belonged to a family of massive evolved stars~[\cite{HD179821}].

The experience obtained in the analysis of the spectra of stars of
different luminosity and mass allows us to hope for a success in the
study of stars with an obscure evolutionary status. The object of the
present study is the star V534\,Lyr (HD\,172324, HIP\,91359), located
at the galactic latitude \mbox{$b=18\fdg58$}. Our interest to this
relatively bright ($V=8\fm16$) star is due to the fact that Trams et
al.~[\cite{Trams1991}] included V534\,Lyr in the list of
high-luminosity stars at high Galactic latitudes (HGL). These authors
attributed the major part of  investigated objects to the post-AGB
candidates. However, due to the lack of data on the IR flux, they
classified V534\,Lyr   only as an alleged candidate in post-AGB. An
additional stimulus to the study of V534\,Lyr for us is the emission
in the  H\,I lines, found in the early paper by Bonsack and
Greenstein~[\cite{Bonsack1956}] and confirmed
later~[\cite{Giridhar2001}]. This star was repeatedly
studied using various photometric
systems~[\cite{Rufener1976,Fernie1983,Hauck1998}]
and spectral methods, but up to now, none of the available
publications has a definite conclusion on its evolutionary status 
(see Table\,\ref{Publ}, in which the basic papers examining  V534\,Lyr are
listed~[\cite{Veen,Giridhar2001,Giridhar2005}]).

\begin{table}[ht!]
\caption{Some previously published V534\,Lyr research results}
\smallskip
\begin{tabular}{l|c|c|c}
\hline
\multicolumn{1}{c|}{Methodology} &  Results & Status & Reference \\
 \hline 
IR Photometry & Lack of IR excess & Post-AGB star?  & [\cite{Veen}] \\
\hline Optical spectroscopy, & $M_V$, $V_r$, & Post-AGB star ? &  [\cite{Giridhar2001}] \\
     method of model atmospheres     & chemical composition       &                                   &  \\
 \hline
Optical Spectroscopy, &  $V_r$ Variability and  & High Speed    \\
method of model atmospheres    &  H\,I profiles,  chemical composition&  HGL star ? & [\cite{Giridhar2005}]    \\
\hline
\end{tabular}
\label{Publ}
\end{table}

The purpose of this paper is to determine the main parameters of V534\,Lyr using the high-quality 
optical spectral data, to search for the possible variability  in time of the spectrum and the pattern of
radial velocities in the atmosphere of the star, and to refine its evolutionary status. 
Section\,\ref{Obs}  briefly describes the used observational data and details of its  processing.
The data on the spectral features obtained from the high-resolution spectra and the detected 
profile features are given in Section\,\ref{results}. We make an attempt here to apply our data 
for the determination of luminosity, remoteness and evolutionary status of the star, as well 
as present a method for determining the parameters of the stellar atmosphere and calculating
the abundances of chemical elements. Section\,\ref{discuss}  discusses the results of our study. 
The main conclusions are summarized  in the final Section\,\ref{conclus}.

\section{Observations and processing  of spectra}\label{Obs}

In this study we used seven high-resolution spectra of V534\,Lyr
($R=60\,000$), obtained with the NES echelle
spectrograph~[\cite{nes2}] of the 6-m BTA telescope on
arbitrary dates in 2010 and 2017, as well as the earliest of our
spectra, which was obtained on July 8, 2000 with the PFES echelle
spectrograph installed in the primary focus of the 6-m telescope with
a lower resolution of \mbox{$R=15\,000$}~[\cite{PFES1,PFES2}],
which imposed some restrictions on its use. Extraction of
one-dimensional spectra from two-dimensional echelle frames was
executed using a modified version~[\cite{Yushkin}] of
the ECHELLE context from the MIDAS software suite. The removal of
traces of cosmic particles was done by the median averaging of two
spectra obtained successively one after another. The wavelength
calibration was carried out using the spectra of the hollow-cathode
Th-Ar lamp. For the subsequent spectrophotometric and positional
reduction of one-dimensional spectra we used a modified version of
the DECH20t code~[\cite{dech}]. The control of
instrumental matching of the spectra of the star and the
hollow-cathode lamp is made from the telluric lines [O\,I], O$_2$ and
H$_2$O.

In more detail, the procedure for measuring the radial velocity $V_r$ based on the
spectra obtained with the NES spectrograph, and sources of error
are described in~[\cite{VR}]. The mean square $V_r$ measurement error
 for stars with narrow absorptions in the spectrum does not exceed
1.0~km\,s$^{-1}$ (one-line accuracy~[\cite{VR}]).
The initial data concerning the obtained spectra  are presented in
the first three columns of Table\,\ref{Velocity}.

\begin{table*}
\caption{The time of observations, registered spectral interval, effective temperature and
 radial velocity values averaged for different groups of spectral lines.  The last two columns give the
 average    $V_r$ values measured from the FeII lines  with high and low lower-level excitation potentials}
\smallskip
\begin{tabular}{c|c|c|c|c|c|c|c}
\hline
Date & UT & $\Delta\lambda$, nm &$T_{\rm eff}$, K&  \multicolumn{4}{c}{$V_r$, km\,s$^{-1}$}   \\
\cline{5-8}
         &    &   &  & emissions& \multicolumn{3}{c}{absorptions} \\
\cline{5-8}
         &  &             &     & Fe\,II   &He\,I+S\,II+Si\,II &  Fe\,II (high)   & Fe\,II (low)  \\  
\hline
   (1)     & (2) &     (3)           &   (4)     & (5)   & (6) & (7) & (8)\\
\hline
 08.07.2000 & 17:23 & 430--780 &       & $-131$  &  $-140$  & $-$145: & $-$143:, $-154^1$ \\  
 06.04.2010 & 01:18 & 516--669 & 10100 & $-131$  &  $-131$  & $-132$  & $-103$ \\  
            &       &          &       &         &          &         & $-152$ \\ 
 01.06.2010 & 23:56 & 522--669 & 10250 & $-123$  & $-120$   & $-120$  & $-107$ \\ 
            &       &          &       &         &          &         & $-156$ \\ 
 30.07.2010 & 23:41 & 443--593 & 10300 &         & $-124$   & $-125$  & $-125$ \\ 
 24.09.2010 & 20:01 & 522--669 & 10800 & $-127$  & $-130$   & $-130$  & $-128$ \\ 
 12.10.2013 & 17:10 & 392--698 & 9750  & $-114$  & $-120$   & $-108$  & $-109$ \\ 
            &       &          &       &         &          &         & $-146$ \\ 
08.06.2017  & 00:01 & 470--778 & 10250 & $-134$  & $-130$   & $-134$  & $-134$ \\  
13.06.2017  & 21:12 & 470--778 &       & $-134$  & $-129$   & $-133$  & $-119^2$, $-113^1$ \\ 
            &       &          &       &         &          &         & $-140^2$, $-152^1$  \\
\hline \multicolumn{8}{l}{{\footnotesize\it $^1$ --the average value
$V_r$ based on Fe\,II\,(42) lines,}} \\ [-3pt]
\multicolumn{8}{l}{{\footnotesize\it $^2$ --the average value $V_r$ based on Fe\,II\, lines
(multiplets 48,\,49,\,74).}} \\ 
\end{tabular}
\label{Velocity}
\end{table*}

\section{Main results}\label{results}

\subsection{Peculiarities of the V534\,Lyr spectrum}

\begin{figure}[ht!]
\includegraphics[angle=0,width=0.9\textwidth]{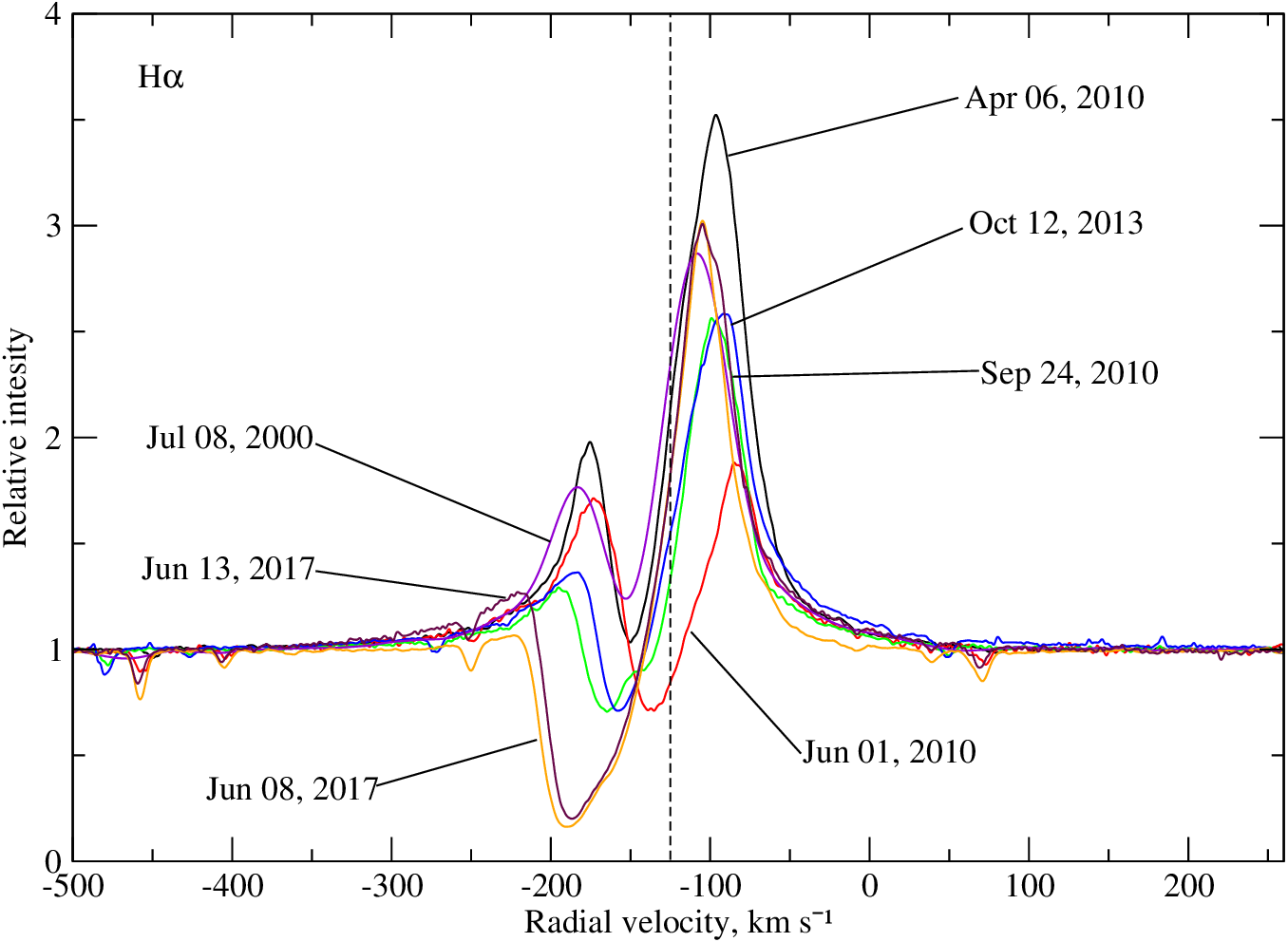}
 \caption{H$\alpha$ line profiles at different observation times. The dashed
   vertical line is the adopted systemic velocity $V_{\rm sys} \approx -125$~km\,s$^{-1}$.} 
\label{Halpha}
\end{figure}

A peculiar feature of the optical spectrum of V534\,Lyr is
a powerful emission of the  H$\alpha$ profile, which was noted
in all the papers dealing with the spectroscopy of this star (see,
for example,~[\cite{Giridhar2001,Giridhar2005}] and references
therein). The H$\alpha$ line, which is present in seven of  eight our
spectra, is the most intense. The variations of its profile with a
two-peak emission  are significant. In the earliest spectrum (July 8,
2000) the entire emission profile is located above the continuum.
However, in the subsequent spectra  the core of the absorption
component which is formed in the surface layers of the atmosphere
falls below the level of the continuum. The position of the
absorption component, as well as the intensities and positions of the
emission components vary, but the intensity of the long-wave peak is
always higher than that of the short-wave peak (Fig.\,\ref{Halpha}).
The shift of the blue border of the absorption component relative to
the vertical dashed line, corresponding to  $V_{\rm sys}$  increases
in the spectrum of September 24, 2010:  \mbox{$V_r -V_{\rm
sys}\approx -50$~km\,s$^{-1}$.}

\begin{figure}[ht!]
\includegraphics[angle=0,width=0.8\textwidth]{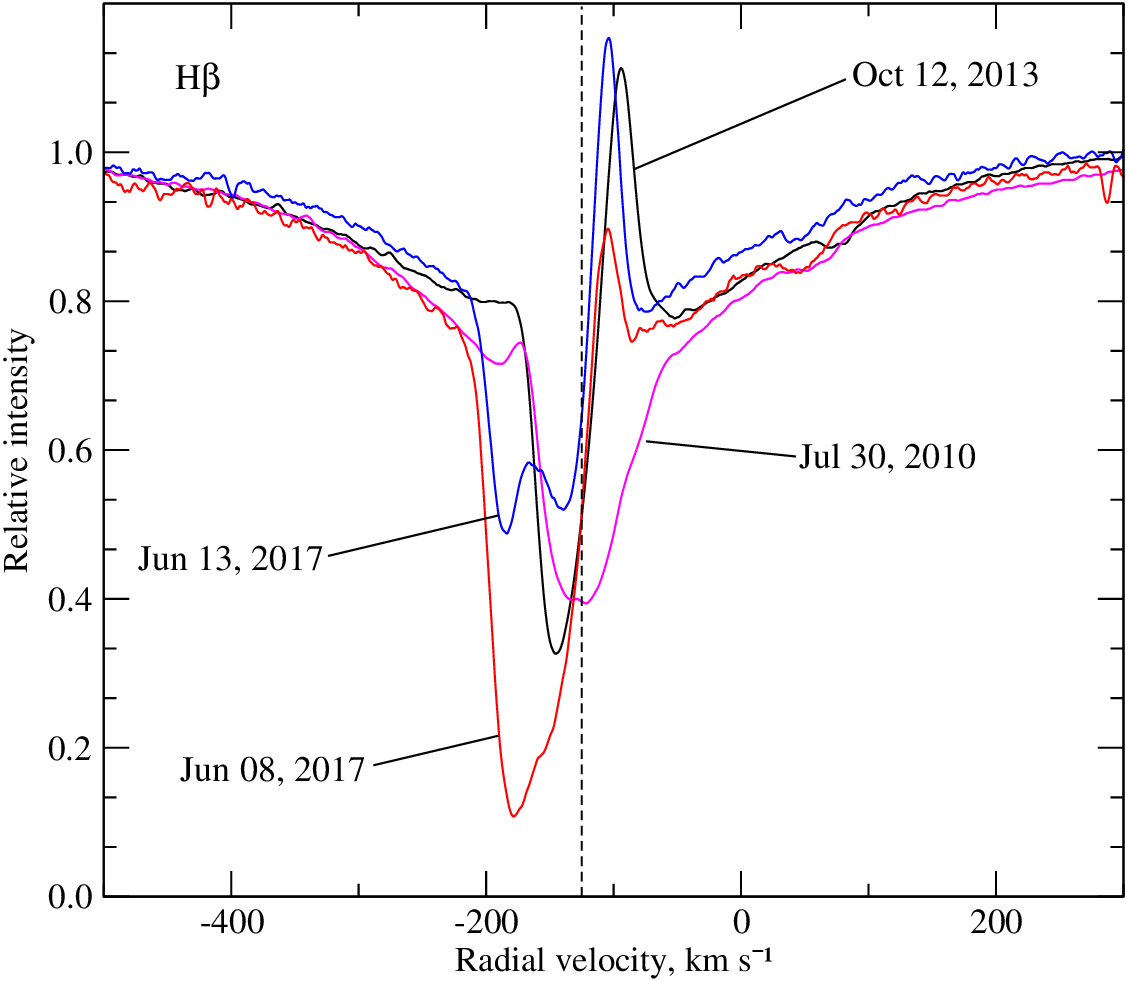}
\caption{The same as in Fig.\ref{Halpha} for the H$\beta$ line.}
\label{Hbeta}
\end{figure}

A drastic profile variation   is observed in the  2017 spectrum.
Instead of the two-peak profile we observe a   P\,Cyg-type profile, the absorption
component of which is shifted by  $V_r - V_{\rm sys}\approx -70$~km\,s$^{-1}$. 
The H$\beta$ profile is registered in a smaller
number of dates, but this is enough to notice how the stronger of the
emission components passes from the red wing to the blue wing
(Fig.\,\ref{Hbeta}). Similar variations of   H$\alpha$
and  H$\beta$ profiles in the spectrum of V534\,Lyr were found in the
paper~[\cite{Giridhar2005}]. As it  can be seen in Fig.\,\ref{Hbeta}, in the 
spectrum obtained on June 8, 2017, there was a change in the H$\beta$  profile type  to the
P\,Cyg-type with the maximum shift of the absorption component. A few
days later, on June 13, 2017, we can see the splitting of the H$\beta$ 
absorption component in the absence of  H$\alpha$ profile variation.

\begin{figure}[ht!]
\includegraphics[angle=0,width=0.9\textwidth,bb=40 200 720 530,clip]{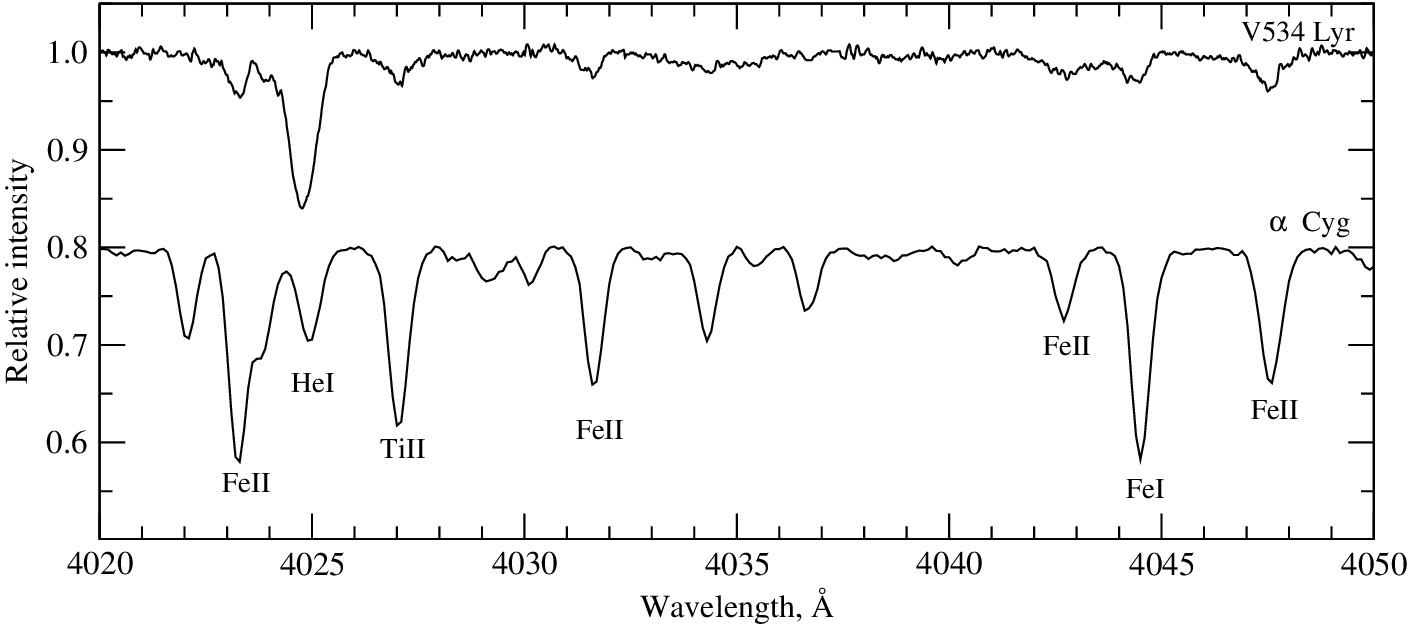}
\includegraphics[angle=0,width=0.9\textwidth,bb=40 200 720 530,clip]{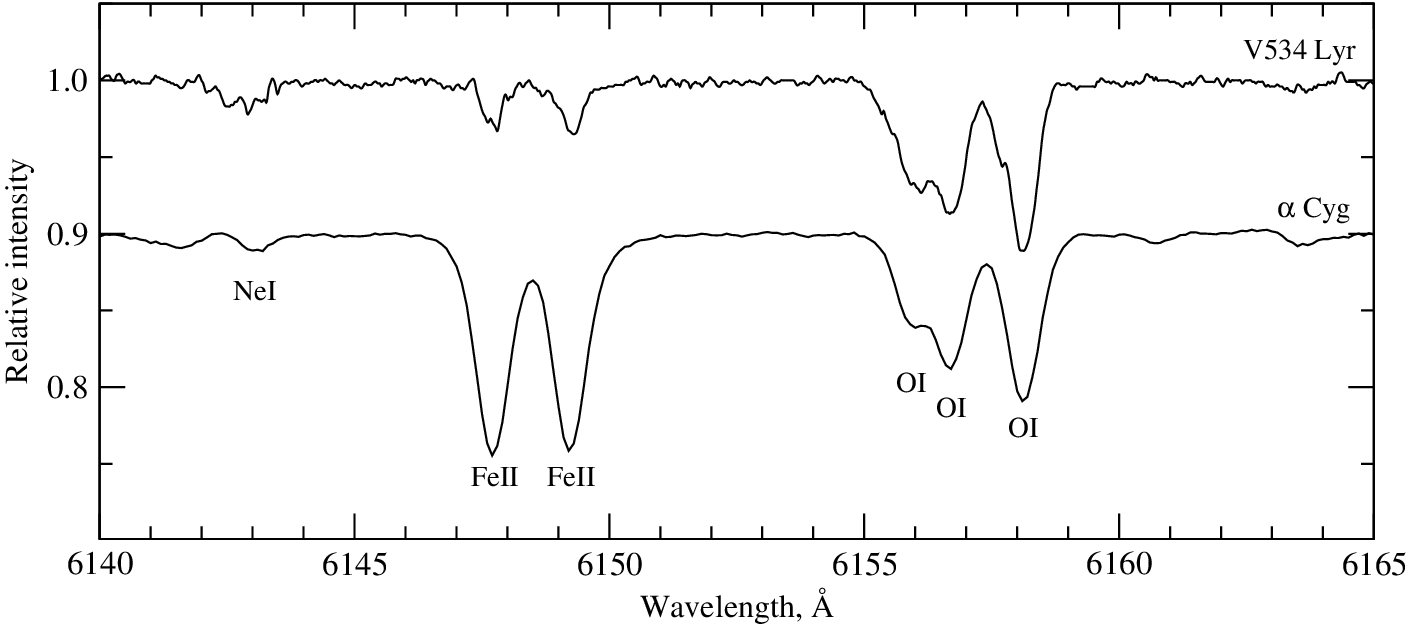}
\caption{A comparison of fragments of the spectrum of V534\,Lyr with similar fragments 
        of the spectrum of the massive supergiant $\alpha$\,Cyg. The main absorptions 
        of the fragment are identified.}
\label{Helium-Oxygen}
\end{figure}

Note that  the V534\,Lyr spectrum  is characterized by a strengthening of neutral 
helium He\,I lines, with an appreciable weakening of   metal lines. This is well illustrated 
in the upper panel of Fig.\,\ref{Helium-Oxygen}, where a fragment of the spectrum of this star 
with the  He\,I\,4026\,\AA{} line is compared with an analogous spectral region of a massive supergiant
$\alpha$\,Cyg. The MK-spectral class of $\alpha$\,Cyg, namely A2\,Ia, is close to V534\,Lyr, 
but the intensities of the  He\,I\,4026\,\AA{} line are markedly different.

\subsection{Radial velocity pattern}

\begin{figure}[t!]
\includegraphics[width=1.0\textwidth,bb=4 20 720 210,clip]{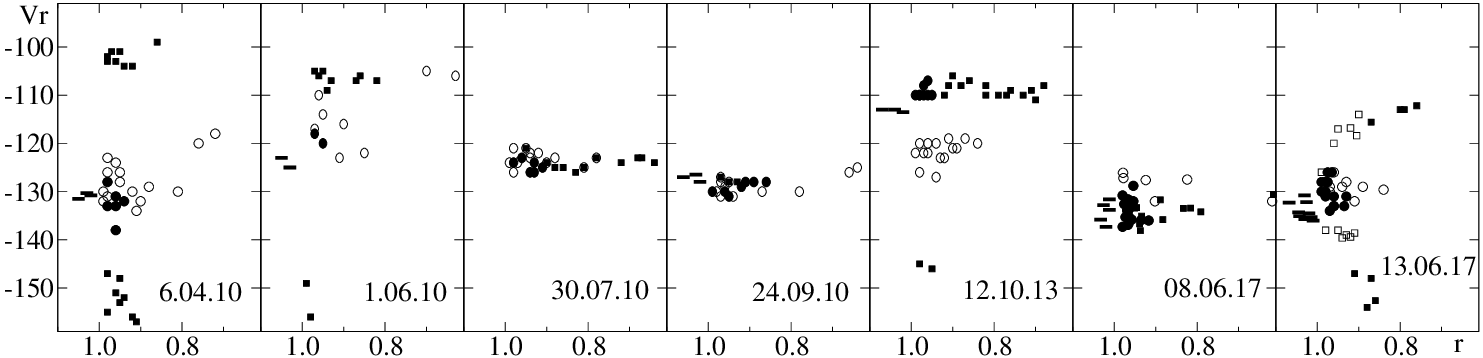}
\caption{The $V_r\,(r)$ dependences, determined from all the obtained spectra. Filled 
         circles -- high-excitation Fe\,II absorptions, squares -- low-excitation Fe\,II 
         absorptions, rings -- He\,I, S\,II, horizontal lines -- Fe\,II\,6318, 6384, 6385\,\AA\ emissions.}
\label{Veloc_var}
\end{figure}

Generalized results of radial velocity measurements for each time of
observations are presented in Table\,\ref{Velocity}. Columns (5)--(8)
contain the values obtained by averaging the velocities for: the
Fe\,II~6318, 6384 and 6385~\AA\ emission, in the 2017 spectra the Fe\,II\,6493, 
7496, 7513\,\AA, emissions, the He\,I, S\,II absorption cores, and the doublet  Si\,II\,(2)
absorption core, the cores of the components of Fe\,II absorptions, respectively. 
The last column lists velocity from the Fe\,II lines with a low lower-level excitation potential. 
In the spectra for four dates, these lines are split into two components, hence  two mean
velocity values are given: based on the long-wave components, and below these values the 
mean velocity for the short-wave components is indicated. 
Let   us stress that for all the times  when it is present in the spectrum 
the splitting reaches large values: $\Delta V_r=20$--$50$~km\,s$^{-1}$. In the 2000 spectrum, 
due to its reduced spectral resolution the absorption component measurements are unreliable. 
For this reason, the data for this spectrum are not given in Fig.\,\ref{Veloc_var}, which presents the
dependences of radial velocity from the central residual intensity of the corresponding line, 
$V_r\,(r)$. One or two signs (in cases of the split absorptions) correspond  to each line. The circles mark the
high-excitation Fe\,II absorptions, the squares -- the low-excitation Fe\,II absorptions, 
the rings -- He\,I, S\,II, the horizontal lines -- mainly the Fe\,II\,6318, 6384, 6385~\AA, emissions, and in
the spectra of 2017 -- the Fe\,II\,6493, 7496, 7513~\AA\ emissions as well. These Fe\,II 
emissions are formed in extended envelopes and, being stationary for some super- and hypergigants, are used to
estimate their systemic velocities $V_{\rm sys}$~[\cite{Chentsov2004}]. According to our data, in
the investigated star V534\,Lyr the velocities determined based on the emissions at different 
dates of 2010 differ by approximately 5~km\,s$^{-1}$, and different lines are shifted by various values,
though all in one direction. In the 2013 spectrum, the position of these emissions is significantly 
different from the previous ones. At the same time, in each of the dates, the position of the emissions is
close to those for the unsplit absorptions, and hence the  velocities in column (5) are close 
to the velocities in column (6), measured from the upper parts of absorptions.

The half-width of the Mg\,II\,4481~\AA{} line in the V534\,Lyr spectrum is $\delta\lambda=0.5$~\AA. 
Using the dependences of the half-widths of this line on the rotation velocity of the star
from~[\cite{Kinman2000}], we get a low rotation velocity $v\sin i=5$--$6$~km\,s$^{-1}$.

\begin{figure}[ht!]
\includegraphics[angle=0,width=0.5\textwidth]{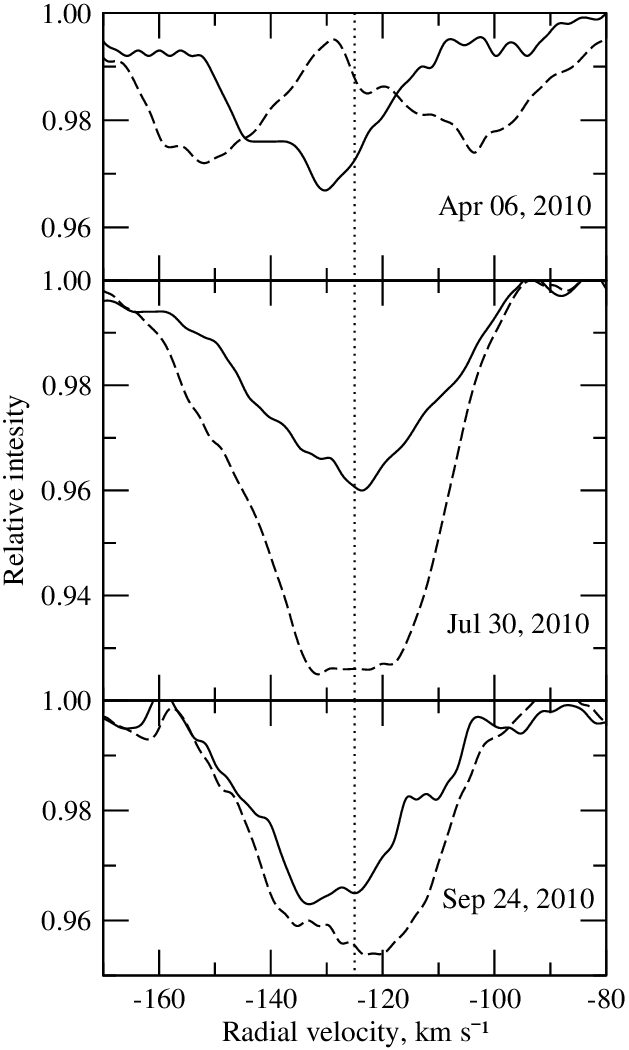}
\caption{The Fe\,II\,5363~\AA (3~eV, dashed) and Fe\,II\,5506~\AA\ (10~eV, solid) line profiles. 
      The vertical dotted line is the same as in Fig.\ref{Halpha} } 
\label{Profiles}
\end{figure}

\begin{figure}[ht!]
\includegraphics[angle=0,width=0.5\textwidth]{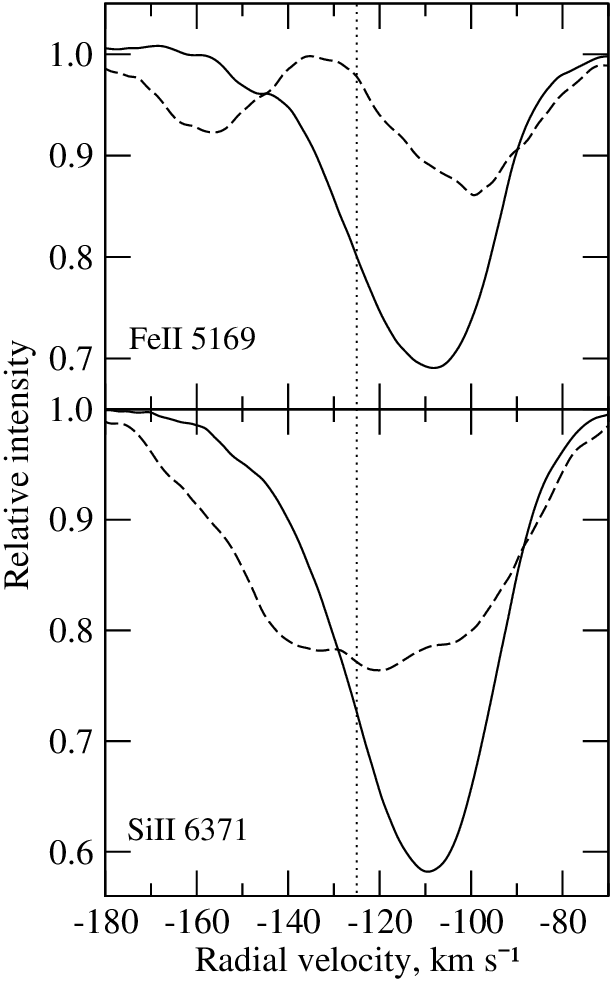}
\caption{The Fe\,II\,5169 and Si\,II\,6371~\AA{} line profiles for two observations: April 6, 2010 (dashed), 
        October 12, 2013 (solid). The vertical dotted line is the same as in Fig.\ref{Halpha}.} 
\label{FeSi}
\end{figure}

Table\,\ref{Velocity} and Fig.\,\ref{Veloc_var} show temporal variations of the positions of all 
lines and profile shapes of some of them. The latter phenomenon is mainly associated with the upper
layers of the atmosphere. This can be seen, for example, from the fact that the low-excitation Fe\,II  
absorptions in the spectra obtained on 6 April, 2010, June 1, 2010 and October 12, 2013 are
split, and the high-excitation absorptions formed deeper remain single in these dates. 
Figure\,\ref{Profiles} demonstrates this for the  Fe\,II\,5363 and 5506~\AA\ lines (the low-level excitation
potentials 3.2~eV and 10.2~eV respectively),  while Fig.\,\ref{FeSi} -- for the Fe\,II\,5169~\AA\ line, 
the strongest of the of the low-excitation absorptions in the visible part of the spectrum 
(its low-level excitation potential is 2.9~eV), and Si\,II 6371~\AA\  (the lower-level potential  is 8.1~eV). 
In the bottom plot of Fig.\,\ref{Profiles} (July 30, 2010~and September 24, 2010) the
profile of Fe\,II\,5363~\AA\ is only slightly asymmetric, and in the top plot (April 6, 2010) 
it is split into two components, close in depth and spaced from each other by 48~km\,s$^{-1}$. The profile of
Fe\,II\,5169~\AA\ (Fig.\,\ref{FeSi}) from April 6, 2010   also consists of two components, though   
possessing different depths, separated by 56~km\,s$^{-1}$. Figure\,\ref{FeSi} shows an intermediate
case of profile transformation. The absorption of Si\,II 6371~\AA\ (the lower-level excitation potential 
of 8.1~eV) in the results of April 6, 2010 observations   is clearly not split, but it is noticeably 
broadened as compared with the data obtained on September 24, 2010 (in the spectrum of July 30, 2010 it is absent).

\begin{table}
\caption{The parameters of some absorptions in the V534\,Lyr spectrum: depths  $R$,  
        widths  $\delta V_r$, km\,s$^{-1}$, and equivalent widths $W_\lambda$, \AA{}}
\begin{tabular}{c|c|c|c}
\hline
Parameter & \multicolumn{3}{c}{Date} \\
\cline{2-4}
        & 6.04.2010 & 30.07.2010 & 24.09.2010\\
\hline
                & \multicolumn{1}{c}{} & \multicolumn{1}{c}{Fe\,II\,5506} \\[5pt]
   $R$          & 0.037     & 0.039      & 0.036 \\
${\delta V_r}$  & 66        & 52         & 45    \\
 $W_{\lambda}$  & 0.030     & 0.026      & 0.022 \\
\hline
                & \multicolumn{1}{c}{} & \multicolumn{1}{c}{Fe\,II\,5363} \\[5pt]
   $R$          & 0.030     & 0.075      & 0.045 \\
${\delta V_r}$  & 85        & 62         & 49    \\
 $W_{\lambda}$  & 0.030     & 0.055      & 0.031 \\
\hline
                & \multicolumn{1}{c}{} & \multicolumn{1}{c}{Fe\,II\,5169} \\[5pt]
   $R$          & 0.140     & 0.370      &       \\
${\delta V_r}$  & 108       & 78         &       \\
 $W_{\lambda}$  & 0.120     & 0.274      &       \\
\hline
                & \multicolumn{1}{c}{} & \multicolumn{1}{c}{Si\,II\,6371} \\[5pt]
   $R$          & 0.240     &            & 0.340 \\
${\delta V_r}$  & 102       &            & 87    \\
 $W_{\lambda}$  & 0.330     &            & 0.340  \\
\hline
\end{tabular}
\label{table3}
\end{table}

Table\,\ref{table3} contains the parameters of the line profiles, presented in Figs.\,\ref{Profiles}
and\,\ref{FeSi}, their depths $R$, widths in the wings (at $r=0.99$) $\delta V_r$ and equivalent widths $W_{\lambda}$.  
In the spectrum of April 6, 2010 the absorptions are shallower and wider in the wings than in the data from 
July 30, 2010 and September 24, 2010, while these differences are relatively small in the
high-excitation Fe\,II  lines, as well as in the  He\,I and S\,II lines, and are noticeably larger in the 
low-excitation Fe\,II lines.
The depths of Fe\,II 5506~\AA, determined from the spectra on the dates indicated are practically identical, 
while for Fe\,II\,5363~\AA\ and Fe\,II\,5169~\AA\ they differ by 2.5 and 2.6~times, respectively. 
The absorption growth of the  Fe\,II\,5363~\AA\ line in the observations of April 6, 2010 was by
September 24, 2010 compensated by the narrowing the way that its equivalent width has returned to the previous value. 
We believe that the broadening and splitting of the low-excitation Fe\,II lines  are related to the features of 
the velocity field in the upper layers of the atmosphere, rather than to the appearance  of the emission
components in them. In the region of the V534\,Lyr spectrum available to us, the latter is confidently 
observed only in the hydrogen lines. At certain moments, as noted above,   the spectrum revealed weak
Fe\,II\,6318, 6384, 6385~\AA\ emissions. In the 2017 spectra, the    Fe\,II emissions in  the longer-wavelength
region of 7388, 7496, 7513~\AA\ were observed.

The difference in velocities is small: the $V_r$  variations   from date to date are from 
11~km\,s$^{-1}$ for He\,I and S\,II to 24~km\,s$^{-1}$ for the high-excitation Fe\,II absorptions. 
From the data given in the tables and   figures with line profiles, it can be concluded that 
the radial velocity gradient in the atmosphere of V534\,Lyr was minimal, and it was the most 
stable on July 30, 2010, September~24, 2010 and on June 8, 2017: the differential shifts of  lines on
these dates are close to the measurement errors, the widths and anomalies of their profile shapes 
are minimal. Assuming that the average radial velocity on these dates were close to the velocity of 
the center of mass of the star, we accept \mbox{$V_{\rm sys} \approx -125$}~km\,s$^{-1}$ 
($V_{\rm lsr}\approx-105$ km\,s$^{-1}$) as the first approximation. This estimate is also close to the average
velocity values   in columns (5)--(7) of  Table\,\ref{Velocity}. Close values of the means and 
amplitudes are given in~[\cite{Bonsack1956}] and~[\cite{Giridhar2005}], but one must bear in
mind that in these papers the velocities were obtained by averaging {\it over all the measured lines}, 
not analyzing their features.

\subsection{Luminosity and distance of the star}

The first estimate of the two-dimensional spectral class of V534\,Lyr was made by V.\,Morgan: B9\,Ib~[\cite{Morgan1950}].
Bonsack and Greenstein~[\cite{Bonsack1956}] changed its spectral class to A0\,Iabe, based on the fact that 
the Balmer  series in the spectrum extends to  H$_{24}$,  and the H$\beta$ and H$\gamma$
lines  have the emission components, while the amplitude of  radial velocity variation is higher 
than the typical for the Ib supergiants. The latter argument reinforces the significant and variable
differential shifts of the lines, in particular, the Balmer progress
observed on October 12, 2013~(the growth of $V_r$ from H$\alpha$ to H$\delta$ from   $-157$~km\,s$^{-1}$ 
from $-133$~km\,s$^{-1}$), and noted by the authors of~[\cite{Giridhar2005,Bonsack1956}] and observed by us
variability of intensities and positions of the emission features of the H\,I line profiles.
 As can be seen in Fig.\,\ref{Halpha}, in the 2000 and 2010 spectra the H$\alpha$ emission  is 
two-peak, and the intensity of the red peak is always above the blue one. However, in June 2017, H$\alpha$
has a P\,Cyg-type profile, which in  a couple of months acquires the character of a two-peak emission. 
The variability of the  H$\beta$ profile is similar: as illustrated in Fig.\,\ref{Hbeta}, its emission
component passes from the red wing to the blue one.
According to the results of observations from June 8, 2017, the  H$\beta$ profile is of a
P\,Cyg-type with a powerful and shifted into the short-wave region absorption, which gets split in 
the spectrum obtained  a few days later, on June 13, 2017.

\begin{figure}[t!]
\includegraphics[angle=0,width=0.45\textwidth]{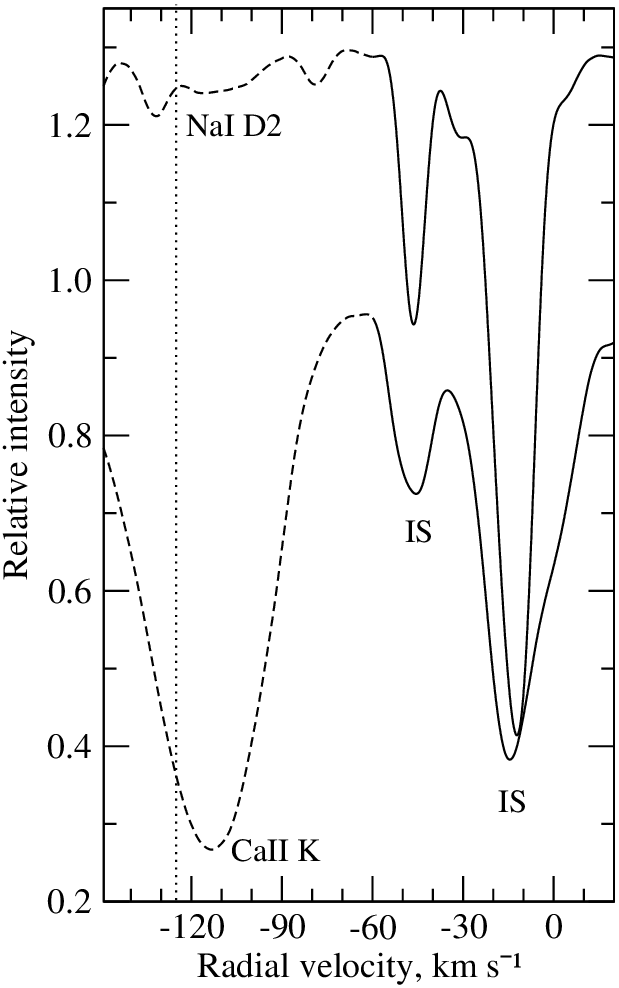}
\caption{The stellar (dashed line) and interstellar parts of the Na\,I\,D2 and Ca\,II\,K line
       profiles in the spectrum obtained on October 12, 2013. The vertical dotted line is as in
      Fig.\ref{Halpha}.} 
\label{NaCa}
\end{figure}

The spectral class A0\,Iab corresponds to the distance to V534\,Lyr of $d=5.7$~kpc~[\cite{Bonsack1956}]. 
The interstellar Na\,I\,(1) and Ca\,II\,(1) line profiles in the stellar spectrum
also imply its great remoteness. The presence in them of the components with \mbox{$V_r=-46$}~km\,s$^{-1}$
(Fig.\,\ref{NaCa}), taking into account the data of Brand and Blitz~[\cite{BrandBlitz}] points to
$d>7$~kpc. A great remoteness of the star could be confirmed by its parallax \mbox{$\pi=0.379$}~mas, 
measured by the GAIA, however, in the DR1 catalog it is burdened with a large error of \mbox{$\pm0.378$}~mas. 
A noticeable own motion  \mbox{$3\farcs6\pm 0\farcs8$} in the~\cite{Leeuven} catalog at \mbox{$d\approx 6$}~kpc  
corresponds to the velocity 103~km\,s$^{-1}$,  directed to the plane of the Galaxy. In combination
with \mbox{$V_{\rm lsr} \approx -105$}~km\,s$^{-1}$  this yields the spatial velocity of about 140~km\,s$^{-1}$.

The presence of interstellar components in the Na\,I\,(1) and Ca\,II\,(1) lines forced us to undertake a search 
for possible DIBs. Luna et al.~[\cite{Luna2008}] in their study of interstellar and circumstellar
 absorption for the sample of post-AGB stars,  found  two such features (5780 and 5797~\AA) in
the spectrum of V534\,Lyr. However, their measurements of the DIB positions are widely scattered. In our
spectra, there are no features that could be reliably identified with the known DIBs. The only feature 
that can be identified with the strongest of the known bands,  5780~\AA\  is very weak in the
spectrum of V534\,Lyr -- its depth is only around 0.015.

The obtained distance  estimate of $d\approx 6$~kpc for the high-latitude star with the apparent 
stellar magnitude of $V = 8\fm58$~[\cite{Fernie1983}] leads us to the value of its absolute magnitude: 
$M_V\approx-5\fm3$, which, according to~[\cite{Straizys}], corresponds to the spectral class
of V534\,Lyr. Vickers et al.~[\cite{Vickers15}], modeling the energy distribution in the spectrum stars,  
obtained the distance of \mbox{$d\approx 3.19\pm0.43$}~kpc. This reduced distance estimate   
leads to a decrease in the absolute magnitude, \mbox{$M_V\approx -3\fm94$},  and worsens the agreement of
spectral class and luminosity.

The luminosity of the star can be determined from the equivalent width of the oxygen IR triplet 
O\,I\,7773~\AA{} in the spectrum of V534\,Lyr: \mbox{$W_{\lambda}=1.99$~\AA.} Using the calibration
from~[\cite{Giridhar2003,IRC2}], we get a very high luminosity, $M_V\approx -8\fm0$. However, this estimate is
burdened by considerable errors, since both calibrations were obtained by the authors
of~[{\cite{Giridhar2003,IRC2}] for the population I~supergiants. The star we are investigating,
V534\,Lyr, obviously does not refer to massive supergiants. In addition, its luminosity estimate 
is also distorted by the specific chemical composition of the atmosphere, primarily, by an anomalous
oxygen abundance.

\subsection{Determination of  model atmosphere parameters and  chemical abundances}\label{analyses}

To determine the main parameters of the model:  effective temperature
$T_{\rm eff}$ and surface gravity $\log g$, we used our standard
method~[\cite{ThetaLeo}], which was successfully used in
the studies of spectra of various types of stars. The parameters
$T_{\rm eff}$ and $\log g$ were determined based on the requirement
of the ionization balance, i.e. the equality of iron  abundance,
calculated from the Fe\,I and Fe\,II lines. Microturbulent velocity
$\xi_t$ was also found using the standard method from the condition
of absence of the dependence of iron abundances $\log
\epsilon$\,(Fe\,I, Fe\,II), determined from a set of lines, on their
equivalent widths $W_\lambda$. To account for possible variations,
the parameters $T_{\rm eff}$ and $\log g$ were found for each
available time of observations. The obtained individual temperatures
$T_{\rm eff}$, given in column~(4) of the
Table\,\ref{Velocity}, differ from each other only
slightly. Their differences are within the error limits of
determining this parameter; therefore, to calculate the abundances of
chemical elements, we used the mean value of $T_{\rm eff}=10\,000$~K.
This estimate agrees  with the previous $T_{\rm eff}$ estimate
from~[\cite{Giridhar2005}] perfectly well, which is an
additional indication of the constancy of stellar temperature. Note
that earlier in the paper by Arellano~Ferro et
al.~[\cite{Giridhar2001}] the condition of solar helium
abundance   in the atmosphere was accepted for determining the
temperature of V534\,Lyr which led to a higher value  $T_{\rm
eff}=11\,500$K.

We controlled the identification of features in the spectrum of
V1\,534\,Lyr involving the atlas~[\cite{Deneb}] with high-resolution
spectra of \mbox{A-supergiants}. The content of chemical elements in
the atmosphere of the studied star is calculated from the equivalent
line widths in the LTE approximation. The calculation of the model
atmosphere, the abundance of chemical elements and the calculation of
synthetic spectra are performed using the WIDTH9 code based on the
Kurucz model grid~[\cite{kurucz}], adapted for the OS~Linux
environment~[\cite{synth}]. The excitation potentials and   oscillator
strengths for all the lines, as well as the broadening constants are
taken from the Vienna Atomic Line Database (VALD)~[\cite{vald}].

Taking into account the above-described peculiarities of the
V534\,Lyr spectrum, we calculated its chemical composition from the
equivalent widths of the absorptions measured in the spectrum of June
8, 2017. In the choice of observational material, we were guided by
the registered wavelength range and the absence of absorption
splitting in the spectrum (Fig.\,\ref{Veloc_var} and
Table\,\ref{Velocity}). To increase the number of lines
we partially involved the data on the equivalent widths of unsplit
lines from the 2013 spectrum. The results of the calculation
performed with the parameters $T_{\rm eff}=10\,000$~K, $\log g=2.5$,
$\xi_t=4.0$ are presented in Table\,\ref{species},
where the number of lines used and the root-mean-square error of
determining the element abundances are indicated. In the last column
of the table the relative values of ${\rm  [X/Fe]_\odot}$ are given,
for the calculations of which we used the information about the
element abundances in the photosphere of the Sun according to Asplund
et al.~[\cite{sun}].

\begin{table}
\caption{The abundances of  chemical elements $\log \epsilon$, calculated with the model 
   parameters   $T_{\rm eff}=10\,000$~K, $\log g=2.5$, $\xi_t=4.0$. The error of the mean 
   $\sigma$, the number of lines  $n$   and the relative element abundances ${\rm  [X/Fe]_\odot}$ are indicated.
  The second column lists the  element abundances in the solar photosphere~[\cite{sun}]}
\smallskip
\begin{tabular}{l|c|r@{$\,\pm\,$}l|c|c}
\hline
\multicolumn{1}{c|}{Element,}  & $\log \epsilon_\odot$ &\multicolumn{4}{c}{V534\,Lyr} \\
 \cline{3-6}
 \multicolumn{1}{c|}{ion}      &[\cite{sun}] & $\log \epsilon$ & $\sigma$ & $n$ & ${\rm [X/Fe]_\odot}$ \\
\hline
He\,I  &10.32 &11.31 & 0.17 & 10 & $+1.27$ \\
C\,II  &8.70  & 8.70 & 0.06 & 3  & $+0.28$  \\
N\,I   &7.83  & 8.65 & 0.21 & 2  & $+1.10$  \\
N\,II  &      & 8.53 & 0.10 & 3  & $+0.98$ \\
O\,I   &8.69  & 8.94 & 0.04 & 18 & $+0.53$  \\
Ne\,I  &7.93  & 8.37 & 0.08 & 6  & $+0.72$ \\
Mg\,I  &7.60  & 7.43 & 0.08 & 5  & $+0.11$ \\
Mg\,II &      & 7.13 & 0.09 & 7  & $-0.19$ \\
Al\,II &6.45  & 6.79 & 0.08 & 2  & $+0.52$ \\
Si\,II &7.51  & 7.53 & 0.16 & 6  & $+0.30$ \\
P\,II  &5.41  & 6.13 & 0.03 & 4  & $+0.90$ \\
S\,II  &7.12  & 7.72 & 0.08 & 9  & $+0.88$ \\
Sc\,II &3.15  & 3.14 & 0.18 & 2  & $+0.27$ \\
Ti\,II &4.95  & 4.61 & 0.05 &20  & $-0.06$\\
V\,II  &3.93  & 3.66 & 0.06 & 2  & $+0.01$  \\
Cr\,II &5.64  & 5.14 & 0.05 & 20 & $-0.28$ \\
Mn\,II &5.43  & 5.69 & 0.10 & 4  & $+0.44$ \\
Fe\,I  &7.50  & 7.23 & 0.05 & 5  & $+0.01$ \\
Fe\,II &      & 7.20 & 0.04 & 66 & $-0.02$ \\
Ni\,II &6.22  & 6.02 & 0.15 & 5  & $+0.08$  \\
Sr\,II &2.87  & 2.50 & 0.01 & 2  & $-0.09$ \\
\hline
\end{tabular}
\label{species}
\end{table}

\section{Discussion}\label{discuss}

The presence of an emission-absorption profile of the  H$\alpha$
line, in which the position of the emission components and their
intensity ratio vary with time is a sign of passage of a shock wave
in the stellar atmosphere. In addition, we see a low-amplitude
variability of radial velocity from the lines with a high excitation
potential, which are formed in deep layers of the stellar atmosphere.
Therefore, we have registered the signs of pulsational instability of
the star. In the spectra of V534\,Lyr we found a previously unknown
feature: splitting of profiles of selected metal absorptions at
separate times of observations. This kind of metal line splitting
indicates the pulsations of W\,Vir-type stars. A good
  example is the velocity field in the atmosphere of
W\,Vir itself~[\cite{Kovtyukh}]. An additional argument confirming the
presence of pulsations is given to us by the analysis of individual
velocities based on the split lines in the spectrum of June 13, 2017,
where the degree of splitting of the low-excitation absorptions
varies depending on the depth of absorptions. This difference is
illustrated by the data in the last line of   Table\,\ref{Velocity}
and the panel of  $V_r\,(r)$ dependences in Fig.\,\ref{Veloc_var},
obtained on June 13,~2017.

The parameters of the model  atmosphere and the   chemical element
abundances in the atmosphere of V534\,Lyr we have determined differ
little from the previously published~[\cite{Giridhar2005}]. 
Here we will consider the features of the chemical composition only briefly.
A detailed analysis of the chemical composition of this star
is difficult for a number of reasons: a very limited set of obtained
abundances of chemical elements; application of the standard model
atmosphere in the case of a star whose atmosphere is unstable and
probably  subject to the influence of shock waves. A contribution
of error due to the neglect of the effect of deviation from the local
thermodynamic equilibrium is also possible. However, their impact on the
metallicity estimate for a hot star is small~[\cite{Cohen}].

The main groups of elements whose relative abundances allow us to
judge on the stage of evolution of the star and its belonging to this
or that population of the Galaxy are as follows: the
\mbox{CNO-triad}, the iron group (Cr, V, Fe, Ni), the
$\alpha$-process light metals (Mg, Al, Si, P, S) and the $s$-process
heavy metals. The abundance of iron in the atmosphere of V534\,Lyr is
slightly lowered:
 \mbox{${\rm
[Fe/H]}_\odot=-0.28$}, which, in combination with a high radial
velocity $V_r\approx -125$~km\,s$^{-1}$  points to the fact that
the star belongs to the thick disk of the Galaxy. The abundance of
metals of the iron group (Cr, V, Mn, Ni) is slightly different
of the iron abundance:
 ${\rm [Met/Fe]}_\odot=+0.06$.

To determine the status of a far-evolved star, the principal in the
element prevalence is the abundance of elements of the CNO group. We
have reliably determined a high nitrogen abundance \mbox{${\rm
[N/Fe]}_\odot=+1.10$} from two N\,I lines of low intensity. The
nitrogen excess  in the atmosphere of the supergiant can be the
result of the first mixing, at which the removal of products of the
CN-cycle during the burning of hydrogen in the core is in force. A
large helium excess, ${\rm [He/Fe]}_\odot=+1.27$ is a result of  a
sequence of nuclear reactions in the stellar core and their
subsequent mixing. The excess of oxygen in V534\,Lyr is illustrated
in the bottom panel of Fig.\,\ref{Helium-Oxygen}, where
the fragments of the spectra of V534\,Lyr and $\alpha$\,Cyg
containing the \mbox{O\,I~6155--6157}~\AA\ oxygen lines are  mapped.
We can also clearly  see   here a weakening of the  Fe\,II ion lines
in the spectrum of V534\,Lyr in comparison with the spectrum of
$\alpha$\,Cyg. The abundances of light metals (Mg, Al, Si, P, S),
synthesized owing to the $\alpha$-process are strengthened in the
atmosphere of V534\,Lyr on the average by $[\alpha/{\rm Fe}]=+0.4$.
The strengthening of light metals is consistent with the fact that
the star pertains to the thick disk of the Galaxy, as evidenced by
the results of the extensive sample of thick-disk stars with a close
metallicity ~[\cite{Reddy2006}].

With respect to the observed set of properties, a hot star
HD\,105262, which is located in the Galaxy at a high latitude of
$b=72\fdg47$ and has the \mbox{B9--A0} spectral class can be regarded
as an analog of V534\,Lyr. In the SIMBAD database this star is
cataloged as a post-AGB supergiant. Earlier this
 star, having no IR excess, was considered as a representative of
an earlier stage of evolution, namely, of the horizontal branch
(hereafter -- HB). Klochkova and Panchuk~[\cite{KP1988}],
using the photographic spectra determined the parameters of $T_{\rm
eff}=8500$~K, the surface gravity $\log g = 1.5$, reduced metallicity
\mbox{${\rm [Fe/H]}_\odot=-1.2$} and a detailed chemical composition
of the atmosphere. In their recent work, Giridhar et
al.~[\cite{Giridhar2010}] obtained the same parameters
of the model, though the metallicity is much lower: \mbox{${\rm
[Fe/H]}_\odot\approx-1.9$}. This difference can be explained by a
higher spectral resolution of the spectra
in~[\cite{Giridhar2010}]. For the comparison we make
between V534\,Lyr and HD\,105262, a large excess of nitrogen, found
in the atmosphere of HD\,105262~[\cite{Giridhar2010}] is
important. Both stars have a low rotation velocity, \mbox{$v \sin
i=6$}~km\,s$^{-1}$. For HD\,105262, rotation velocity is measured
from the high-resolution spectra in~[\cite{Martin}].
Belonging to the HB stage  can explain the absence in V534\,Lyr and
HD\,105262 of the circumstellar dust and infrared excess caused by
it.

The  spectral features, similar to the peculiarities of the spectrum
V534\,Lyr are observed in the A-supergiant BD+48$\degr$1220.
The optical high-resolution spectra based on which a significant
variability of the  H\,I and metal line profiles were obtained over
several observational seasons at the 6-m BTA telescope combined with
an echelle spectrograph~[\cite{IBVS}]. The analysis of
this spectral material by the atmospheric model method   with the
parameters \mbox{$T_{\rm eff}=7900$}~K, $\log g=0.0$,
\mbox{$\xi_t=6.0$} has shown that the metallicity   BD+48$\degr$1220  
is close to solar: \mbox{${\rm [Fe/H]}_\odot=-0.10$}~[\cite{IRAS05040}], and the
chemical composition of its atmosphere differs little from that of
V534\,Lyr. As in the case of V534\,Lyr, a large helium excess was
detected, \mbox{${\rm [He/H]}=+1.04$} and an equally significant
excess of oxygen \mbox{${\rm [O/Fe]}_\odot = + 0.72$}~dex. At that,
the carbon excess is small,  \mbox{${\rm [C/Fe]}_\odot=+0.09$}, and
the ratio \mbox{${\rm [C/O]}<1$}. The abundances of light metals are
changed:  \mbox{${\rm [Na/Fe]}_\odot =+0.87$} at \mbox{${\rm
[Mg/Fe]}_\odot =-0.31$}. But the most important -- in the atmosphere
of  BD+48$\degr$1220 a large excess of lithium is
detected\linebreak \mbox{${\rm [Li/Fe]}_\odot=+0.62$}, what indicates
the transfer of this element, synthesized at the AGB stage, into the
atmosphere. In addition, it was concluded on the probable efficiency
of the mechanism of selective separation of chemical elements onto
the dust particles of the envelope. A full set of the available data
(luminosity \mbox{$M_v \approx -5^{\rm m}$}, velocity $V_{\rm lsr}
\approx -20$~km\,s$^{-1}$, reduced metallicity and the features of
the optical spectrum and chemical composition) confirms the status of
the \mbox{O-rich} post-AGB star with the initial mass of
$4\div9 \mathcal{M}_\odot$. Thus, the star BD+48$\degr$1220,  having a
number of properties close to V534\,Lyr is a supergiant at a more
advanced stage, after AGB.

We therefore see that the observed properties of V534\,Lyr, with the
exception of  its high luminosity and location outside the plane of
the Galaxy  do not give any grounds for classifying it as a post-AGB
star. The main fact is that V534\,Lyr does not have an IR excess or
chemical composition anomalies expected for the  post-AGB supergiant.
Let us consider other possible variants of its status. The  lack of a
dust envelope at a sufficiently high luminosity allows to consider it
as an object at the stage after the red giant branch, located at an
evolutionary stage above the HB. The source of energy release in the
stars at this stage, past the helium flash in the core is the burning
of helium in the core and hydrogen in the layer. A large mass loss is
not expected in these supergiants. To confirm this status of the star
one can rely on a convenient diagram $\log g$--$T_{\rm eff}$ (Fig.~1
in~[\cite{Martin}]). The set of fundamental parameters
of V534\,Lyr corresponds to the HB stars in the diagram. A low
rotation velocity \mbox{$v \sin i=5$--$6$}~km\,s$^{-1}$ is also
consistent with the supposed status near the BHB: it is known that
\mbox{HB stars} rotate
slowly~[\cite{Peterson,Cohen}]. The
results we obtained here
  for V534\,Lyr make up another confirmation
of the conclusion~[\cite{Envelopes,Evolution}] on the heterogeneity of the sample of
PPN candidates.

A combination of the observed features of V534\,Lyr: the
pre\-sence of pul\-sations in  deep layers of the atmosphere, reduced
metallicity, the type and variability of the emission-absorption
H$\alpha$ profile allows us to assume that the star belongs to
pulsating
 population II  stars that are located in the instability band above the HB
and evolve to the AGB. Depending on the mass and hence on the
period of pulsations, it can be a  BL\,Her or W\,Vir-type star.
A direct indication of the status of the  pulsating star is provided by
the optical spectrum features: two-peak emissions and time-variable
  H$\alpha$ and H$\beta$ line profiles (Figs.\,\ref{Halpha} and\,\ref{Hbeta}),
the presence  at some times of observations of the absorption
splitting (Fig.\,\ref{Profiles}) and the presence of a velocity
gradient in the atmosphere, registered on June 13, 2017, which is
clearly visible in Fig.\,\ref{Veloc_var}.

It is useful to pay attention here to the study of the chemical
composition of a sample of 19 variable  population II
stars~[\cite{Maas}], based on the high-resolution
spectroscopy. To calculate the abundances of chemical elements, the
authors used only the spectra in which there were no signs of
splitting (or asymmetry) of absorptions or emissions in the Balmer
line series. Maas et al.~[\cite{Maas}] made a conclusion
about the fundamental difference of the chemical composition in
BL\,Her or W\,Vir-type stars: in the atmospheres of the BL\,Her-type
stars, the Na abundance is higher than that in the W\,Vir-type stars.
This  conclusion gives us grounds to more likely  classify the
investigated star V534\,Lyr to the Virginids: we failed to find any
of the Na\,I lines in the spectra of V534\,Lyr, expected in the case
of an excess of this element in the atmosphere.

Typical representatives  of the  W\,Vir-type stars are the
globular cluster members. An example is the Virginid V1~(K\,307), a member of the
globular cluster M\,12. The authors of~[\cite{M12}] determined its
main parameters  (\mbox{$T_{\rm eff}=5600$}~K, $\log
g=1.3$)  and  metallicity ${\rm [Fe/H]}=-1.27$,   which agrees with
the metallicity of other members of M\,12. The atmosphere of this Virginid
revealed altered abundances of CNO-elements, while nitrogen is in
excess: ${\rm [N/Fe]}=+1.15$~dex.
 Relative abundances of Na and the \mbox{$\alpha$-process}
elements:  Mg, Al, Si, Ca, and Ti are also strengthened in a various
degree. The main feature of the spectrum of K\,307 is the presence of
complex absorption-emission profiles of the H$\beta$ and H$\alpha$
lines (see Fig.~2 in~[\cite{M12}]), similar to the
profiles in the spectrum of V534\,Lyr.

The basic kinematic and chemical parameters for an extensive
 sample of HB stars in the thick disk are considered by Kinman et al.~[\cite{Kinman2009}].
 Our assumption that V534\,Lyr
can be a thick disk star in the stage above HB (\mbox{post-HB}) is confirmed by a large
spatial velocity
of the studied star, the absence of a dusty circumstellar envelope, and a
specific chemical composition of the atmosphere. A  factor complicating
the fixation of the stellar status is that stars of several evolutionary stages
can be located above the HB: it can be
young enough stars, evolving to RGB, as well as the stars more
 advanced in the stellar evolution after or yet before the AGB.

The answer to the question critical for more specific conclusions is
on the possible binarity of V534\,Lyr. The $V_r$ variations with time
are probably caused by pulsations, however, a spectral SB\,1-type
binarity is not yet excluded. The found velocity variability based on
the Fe\,II emission (\ref{Velocity}) can be considered
an indication on the possible binarity of the star. Note here the
results of {\c S}ahin and Jeffery~[\cite{S2007}], who,
having analyzed the variability of photometric data for V534\,Lyr did
not find an explicit period and assumed that the variations can be
short-period. There are currently no reliable photometric and
spectral data required for the study and determination of the
variability parameters  of brightness and spectrum. Therefore, it is
extremely necessary to conduct a long-term and detailed monitoring of
V534\,Lyr for the final conclusions about the nature of its
variability.

\section{Conclusions}\label{conclus}

Based on the observations at  the 6-m telescope with the NES echelle spectrograph
 ($R$\,=\,60\,000)  we  investigated the features of the optical
spectrum of V534\,Lyr, a high-latitude star with an uncertain status.
For all the observation dates, heliocentric radial velocities $V_r$
correspond to the positions of all the components of metal
absorption, as well as the Na\,I D and H$\alpha$ lines were measured.
The analysis of the velocity field from the lines of different nature
revealed a small-amplitude $V_r$ variability  from the lines with a
high excitation potential, which are formed in   deep layers of
stellar atmosphere, and allowed us to estimate the systemic velocity
\mbox{$V_{\rm sys} \approx -125$~km\,s$^{-1}$ ($V_{\rm
lsr}\approx-105$~km\,s$^{-1}$).}  The estimated distance
\mbox{$d\approx 6$}~kpc for a high-latitude star leads to the
 absolute magnitude estimate of \mbox{$M_V\approx -5.3^{\rm m}$}, which agrees with
the spectral classification of V534\,Lyr.

We discovered an as yet unknown for this star spectral phenomenon:
splitting of profiles of the selected metal absorptions at separate
times of observation. For all the moments when splitting is present
in the spectrum, it reaches large values: \mbox{$\Delta
V_r=20$--$50$~km\,s$^{-1}$.}

The spectral class of the star is close to A0\,Ib, while its effective
temperature is  $T_{\rm eff}\approx10\,500$~K. Metallicity, reliably
determined by metal lines of the iron group (Cr, V, Mn, Ni)
slightly differs from the abundance of this element:  ${\rm
[Met/Fe]}=+0.06$.  The nitrogen and helium excess indicates
an advanced evolutionary stage of the star. A low
 iron-group metal abundance in combination with a high radial velocity
indicates that the star belongs to the thick disk of the Galaxy.

The whole of observed peculiarities of    V534\,Lyr: the presence of
probable pulsations in   deep layers of the atmosphere, splitting of
metal absorption profiles  with the low lower-level excitation
potential registered at separate times of observations, reduced
metallicity, type and variability of the emission-absorption
H$\alpha$ and H$\beta$ profile allows us to place the star amid the
pulsating population II  stars, which are located in the instability
band near the HB. In general, we can conclude that there is a
complete discrepancy between the properties of V534\,Lyr we have
registered and the post-AGB stage, where the star was referred  to in
the previously published papers.

\section*{Acknowledgements}
The work was supported by the Russian Science Foundation (grant No.~14--50--00043, 
direction ``Magnetometry of Stars'' ). E.G.S. thanks
the Russian Foundation for Basic Research for the financial support
(project 16--02--00587\,a). This research has made use of the SIMBAD
database, operated at CDS, Strasbourg, France, and NASA's
Astrophysics Data System.


\begin{flushright}
{\it Translated by A.~Zyazeva}
\end{flushright}

\end{document}